\documentclass[journal,article,accept,moreauthors,dvipdfm,12pt,a4paper]{mdpi}
\lastpage{x}
\doinum{10.3390/------}
\pubvolume{xx}
\pubyear{2014}
\history{Received: xx / Accepted: xx / Published: xx}

\Title{Tsallis Distribution Decorated With Log-Periodic
Oscillation}

\Author{Grzegorz Wilk$^{1,}$*, Zbigniew W\l odarczyk$^{2}$}

\address{%
$^{1}$ National Centre for Nuclear Research,~Department of
Fundamental Research, Ho\.za 69, 00-681; Warsaw, Poland\\
$^{2}$ Institute of Physics, Jan Kochanowski University,
\'Swi\c{e}tokrzyska 15; 25-406 Kielce, Poland}

\corres{wilk@fuw.edu.pl, +48-22-621 60 85}

\abstract{ In many situations, in all branches of physics, one
encounters power-like behavior of some variables which are best
described by a Tsallis distribution characterized by a
nonextensivity parameter $q$ and scale parameter $T$. However,
there exist experimental results which can be described only by a
Tsallis distributions  which are additionally decorated by some
log-periodic oscillating factor.  We argue that such a factor can
originate from allowing for a complex nonextensivity parameter
$q$. The possible information conveyed by such an approach (like
the occurrence of complex heat capacity, the notion of complex
probability or complex multiplicative noise) will  also be
discussed.}


\begin{document}

\section{Introduction}
\label{sec:I}

In many situations, in all branches of physics, one encounters
behavior of some variables $X$ which become pure power
distributions for large values of $X$ and exponential for $X
\rightarrow 0$.  Because of this they are known as power-like
distributions and in many cases they are identified with a Tsallis
distribution \cite{Tsallis},
\begin{equation}
F(X)=A~\left[1-\left(1-q\right)\frac{X}{T}\right]^{1/(1-q)},
\label{T}
\end{equation}
characterized by a scale parameter $T$ and parameter $q$ known as
nonextensivity parameter ($A$ is normalization)\footnote{The
reason being the fact that Eq. (\ref{T}) is also emerging
from nonextensive statistical mechanics \cite{Tsallis}.}.
Obviously, for $X \rightarrow 0$ distribution (\ref{T}) becomes
the usual Boltzmann-Gibbs exponential formula with temperature
$T$, but it becomes pure exponential (i.e., BG) also for $q
\rightarrow 1$. For $q \neq 1$ and large values of $X$ it becomes
pure power distribution not sensitive to scale parameter $T$.

To fully recognize the nontrivial character of distribution
(\ref{T}), one must realize that, usually, in different parts of
phase space of the variable $X$, one encounters (or, rather, one
expects) a dominance of different (if not completely disparate)
dynamical factors. This is best seen in the processes of
multiparticle production at high energies (the best known to us).
They  will serve here to exemplify our further consideration concerning
some specific log-periodic oscillations, apparently visible in such
processes,  which must be therefore somehow hidden in the
original distribution (\ref{T}).

Before proceeding further, we shall briefly summarize the present
status of application of Tsallis distributions in this context,
concentrating only on multiparticle production processes.
They comprise of many different mechanisms in
different parts of phase space. Limiting ourselves only to
particle production in the central rapidity region and to
distribution of their transverse momenta $p_T$, it is customary to
divide this production into independent {\it soft} and {\it hard}
processes populating different parts of the transverse momentum
space\footnote{A few words of definition concerning this phase
space is necessary. A produced particle has some momentum $\vec{p} =
\left[ p_L, \vec{p}_T \right]$. Its longitudinal part, $p_L$, is
defined as parallel to the axis of collision, its transverse part,
$\vec{p}_T$ as perpendicular to that axis. They are defined by
means of rapidity $y$ variable, $y = \frac{1}{2}\ln\frac{E +
p_L}{E-p_L}$, as, respectively, $p= |\vec{p}| = m\sinh y$, whereas
energy of particle, $E = \sqrt{m^2 + p^2} = m \cosh y$. Central
rapidity means $y = 0$. In what follows, our $X$ from Eq.
(\ref{T}) will be identified with transverse momentum, $X=p_T$.}
separated by a momentum scale $p_0$. As a rule of thumb, the
spectra of the soft processes in the low-$p_T$ region are (almost)
exponential, $F (p_T )$$\sim$$\exp (-p_T/T)$, and are usually
associated with the thermodynamical description of the hadronizing
system. The $p_T$ spectra of the hard process in the high-$p_T$
region are regarded as essentially power-like, $F (p_T
)$$\sim$$p_T^{-n}$, and are usually associated with the hard
scattering process (for relevant literature concerning both parts
see \cite{ISMD2014}). However, it was very soon recognized that
both descriptions could be replaced by a simple interpolating formula
\cite{Michael},
\begin{eqnarray}
F(p_T)=A \left ( 1 + \frac{p_T}{p_0} \right ) ^ {-n}, \label{CM-H}
\end{eqnarray}
that becomes power-like for high $p_T$ and exponential-like for
low $p_T$. The reasoning was that for high $p_T$, where we are
usually neglecting the constant term, the scale parameter $p_0$
becomes irrelevant, whereas for low $p_T$ it becomes, together
with the power index $n$, an effective temperature $T = p_0/n$. The same
formula re-emerged later to become known as the {\it QCD-based
Hagedorn formula} \cite{H}. It was used for the first time in
\cite{UA1} and became one of the standard phenomenological
formulas for $p_T$ data analysis
\cite{PHENIX,STAR,CMS,ATLAS,ALICE}. In the mean time it was
realized that both formulas are, in fact, identical once
\begin{equation}
n = \frac{1}{q - 1}\quad {\rm and}\quad p_0 = nT ,\label{qTn}
\end{equation}
and therefore they can be used interchangeably\footnote{Both Eqs.
(\ref{T})) and (\ref{CM-H}) have been widely used in the
phenomenological analysis of multiparticle productions, including
situations where the nowadays observed spectra extend over many
orders of magnitude,
\cite{BCM,Beck,RWW,qWW1,qWW,PHENIX,STAR,CMS,ATLAS,ALICE,Wibig,Biro,BiroC,JCleymans,ADeppman,Others,WalRaf,CYW}.
Up to now such possibility of testing Tsallis distribution offered
only cosmic ray fluxes, cf. \cite{qCR}.}

This distribution is usually used in a thermodynamical content in
which the scale parameter $T$ is identified with the usual
temperature (although such identification cannot be solid
\cite{RGC}) and with a real power index $n =1/(q-1)$ (or real
nonextensivity parameter $q$).  Actually, a Tsallis distribution
can be regarded as a generalization to real power $n$ (or $q$) of
such well known distributions as the Snedecor distribution (with
$n = (\nu + 2)/2 $ and integer $\nu$, which for $\nu \rightarrow
\infty$ it becomes exponential distribution).

In \cite{LPOWW} we investigate the case when $q$ is a complex
number. We shall review our results in this field in the next
section adding examples where log-periodic oscillations occur at
different energies and for different collision systems. In Section
\ref{sec:III} we discuss the possible consequences of complex
nonextensivity parameter including some new recent developments in
this field (as complex probability and complex multiplicative
noise). The final section contains our conclusions and summary.

\section{Log-periodic oscillations in Tsallis distribution - complex power index}
\label{sec:II}

\begin{figure} [h]
\begin{center}
\includegraphics[scale=0.45]{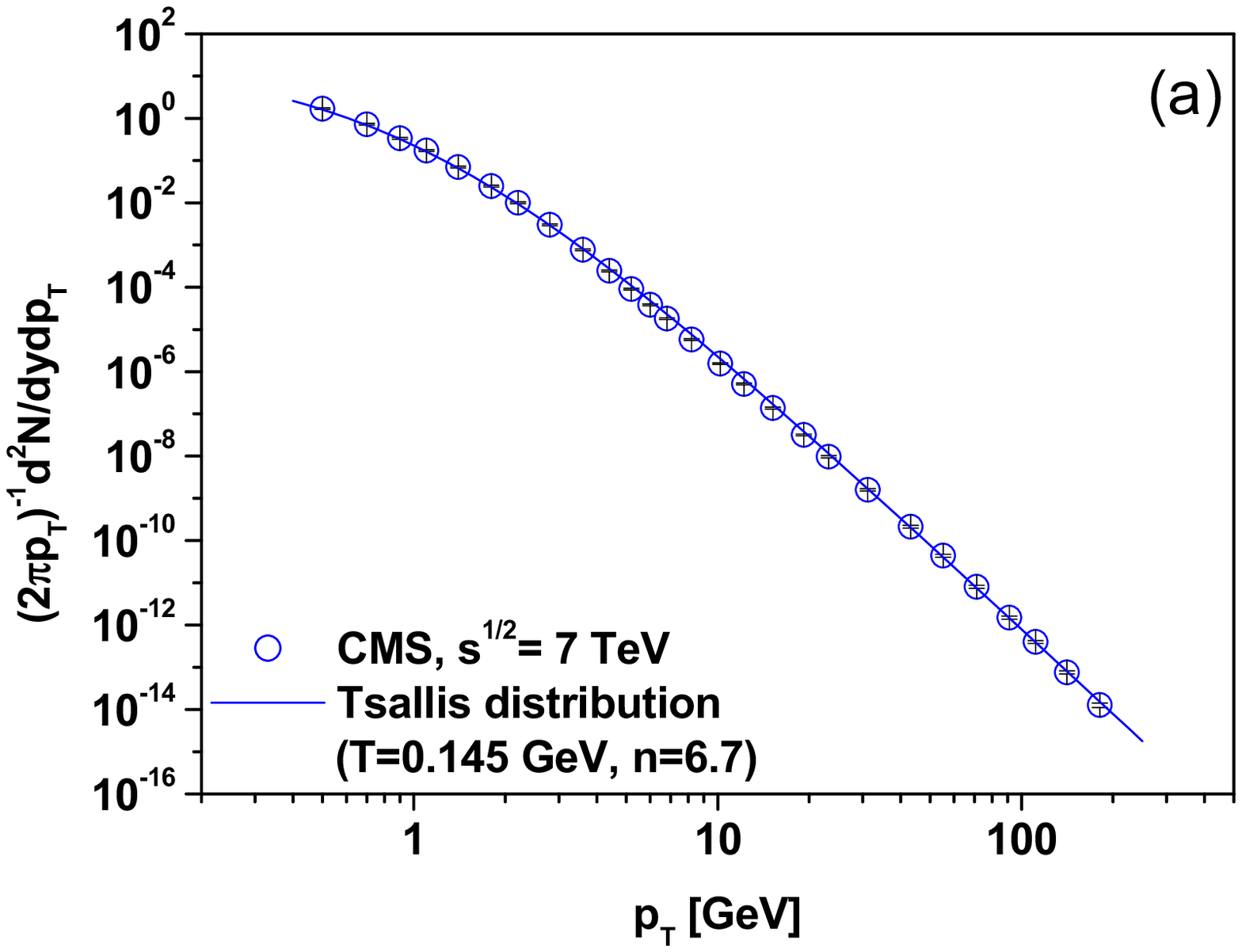}
\includegraphics[scale=0.45]{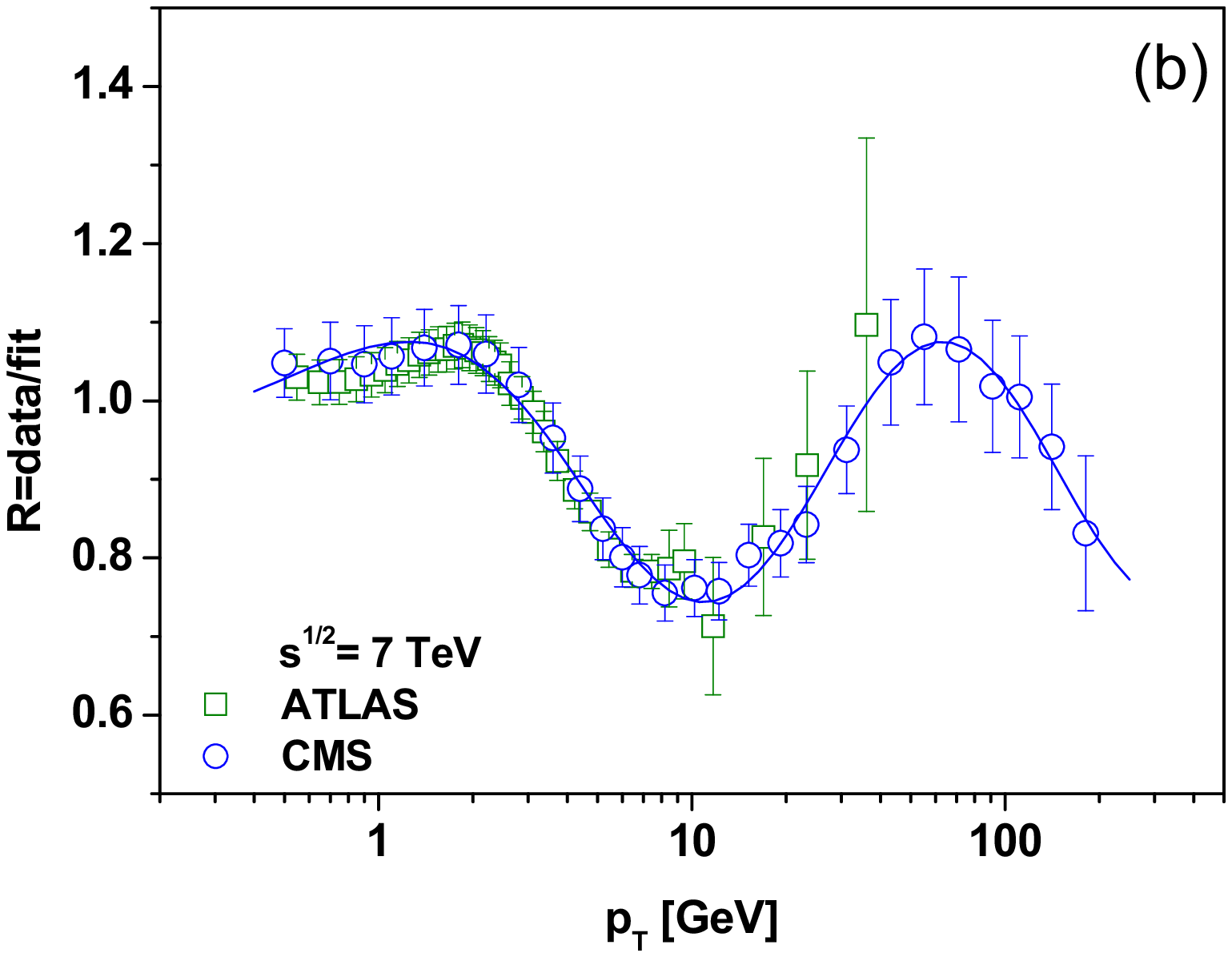}
\includegraphics[scale=0.45]{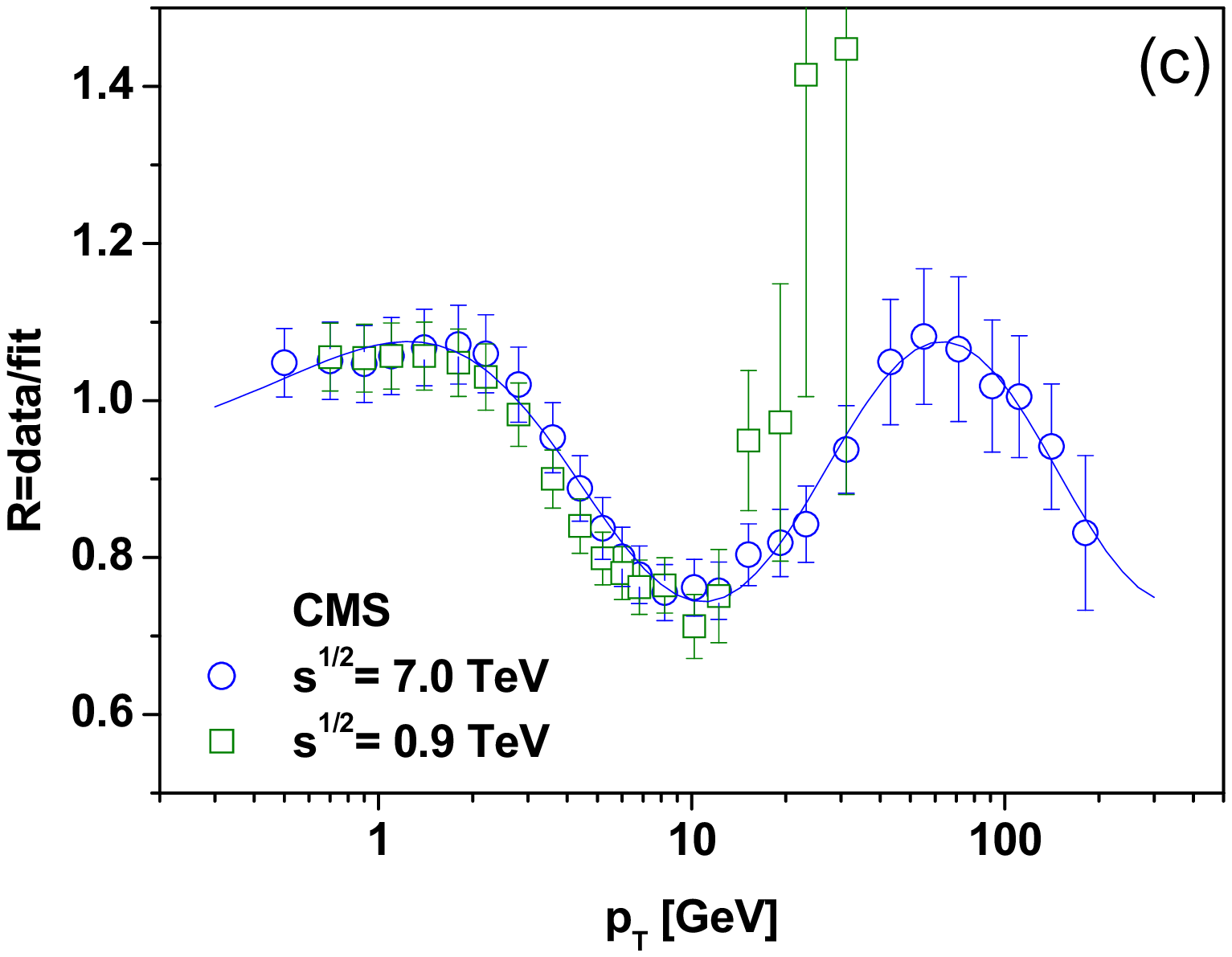}
\includegraphics[scale=0.45]{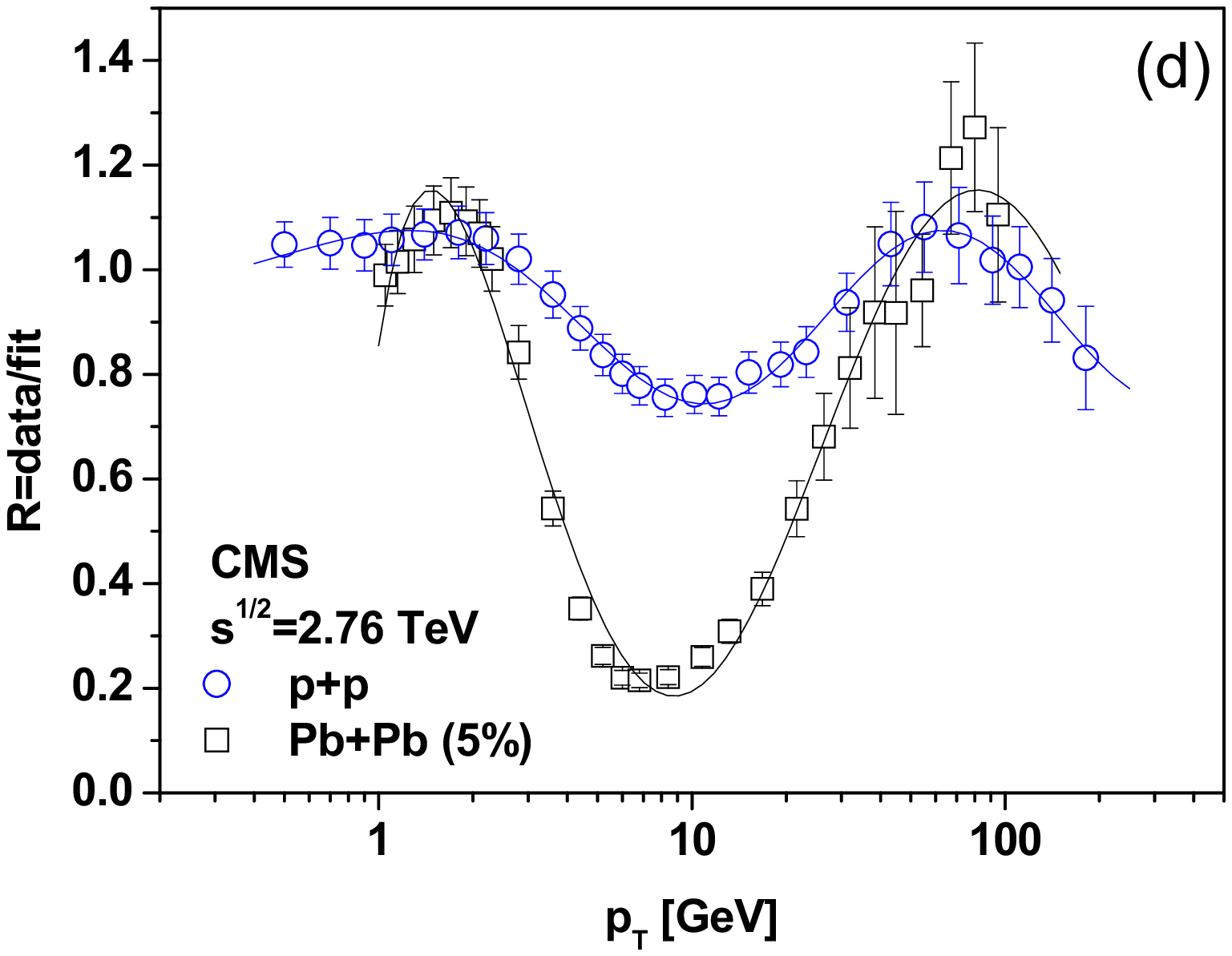}
\end{center}
\caption{(Color online) Examples of log-periodic oscillations.
$(a)$ $dN/dp_T$ for highest energy $7$ TeV, the Tsallis behavior
is evident. Only CMS data are shown \cite{CMS}, others behave
essentially in an identical manner. $(b)$ Log-periodic
oscillations showing up in different experimental data like CMS
\cite{CMS} or ATLAS \cite{ATLAS} taken at $7$ TeV. $(c)$ Results
from CMS \cite{CMS} for different energies. $(d)$ Results for
different systems ($p+p$ collisions compared with $Pb+Pb$ taken
for $5$ \% centrality \cite{CMSPb}. Results from ALICE
\cite{ALICEPb} are very similar. Fits for $p+p$ collision at $7,
~2.76$ and $0.9$ TeV are performed with  $q = 1.139 + i\cdot
0.0385,~~1.134 + i\cdot 0.0269$ and $1.117 + i\cdot 0.0307$,
respectively. Fit for central $Pb+Pb$ collisions at $2.76$ TeV is
done with $q = 1.135 + i\cdot 0.0321$. See text for more details.}
\label{Summary}
\end{figure}

Recently, the experiments \cite{CMS,ATLAS,ALICE} at the Large
Hadron Collider (LHC) at CERN provided new data in a very large domain of
transverse momenta, $p_T$, phase space. They turned out to be
extremely interesting because of the following:
\begin{itemize}
\item They allow us to test the standard Tsallis formula, Eq. (\ref{T}),
over $\sim 14$ orders of magnitude.
As can be seen in Fig. \ref{Summary}a, the observed $p_T$ distributions
of secondaries produced in proton-proton collisions in these experiments can be very well
reproduced (cf. also \cite{CYW}\footnote{These secondaries were produced at midrapidity, i.e.,
for $y \simeq 0$ for which, for large transverse momentum, $p_T > m$ (where $m$ is the mass of the
particle), one has that, approximately, the energy of particle, $E
\simeq p_T$, i.e., it practically coincides with $p_T$.}.
\item And, what is of special importance to us, they disclose some features
which suggest a departures from the single form of Eq. (\ref{T}), cf.Figs.
\ref{Summary} b-c.. Apparently they could not be seen in previous experiments
because they seem to be connected with rather large values of transverse momenta,
not available earlier..
\end{itemize}
However, whereas fits to Eq. (\ref{T}) look pretty good, closer inspection
shows that the ratio of {\it data/fit} is not flat. It shows some kind of visible
oscillations, cf. Fig. \ref{Summary}b. These are the oscillations we have
mentioned before.

It turns out that these oscillations cannot be compensated, or erased,
by any reasonable change of fitting parameters. Moreover, they are
visible by all three experiments CMS, ATLAS, ALICE. The only condition
for such an effect to be visible is that the experiment covers a sufficiently large
domain of transverse momenta $p_T$, cf. Fig. \ref{Summary}b. It is
also seen at all energies covered by these experiments, cf. Fig.
\ref{Summary}c. And, finally, as Fig. \ref{Summary}d shows, this
effect is also visible (and is even more pronounced) in nuclear
collisions. When taken seriously, it turns out that to account for
these oscillations one has to "decorate" distribution $f\left( p_T
\right)$ from Eq. (\ref{T}) (i.e., one has to multiply it) with some
log-periodic oscillating factor. It is is usually taken in the
form \cite{Scaling}:
\begin{equation}
R(E)= a + b\cos\left[ c\ln(E + d) + f\right]. \label{eq:Factor}
\end{equation}

Before proceeding any further let us remember that such log-periodic
oscillations are widely know in all situations in which one
encounters power distributions.  In fact, such behavior has been
found in earthquakes \cite{EQ}, escape probabilities in chaotic
maps close to crisis \cite{CHM}, biased diffusion of tracers on
random systems \cite{BD}, kinetic and dynamic processes on random
quenched and fractal media \cite{RQFM}, when considering the specific heat
associated with self-similar \cite{VMdST} or fractal spectra
\cite{TdSMVM}, diffusion-limited-aggregate clusters \cite{DLM},
growth models \cite{GM}, or stock markets near financial crashes
\cite{FM}, to name only a few examples. However, in all these
cases the basic distributions were scale free power law,
without any scale parameter (here $T$) and without a constant term
governing their $X < nT$ behavior.

In the context of nonextensive statistical mechanics log-periodic
oscillations have first been observed and discussed while analyzing
the convergence dynamics of $z$-logistic maps \cite{UT}. In this paper
we shall propose another way of introducing such
oscillations to Tsallis distributions. It will be based on
allowing the power index $n$ (or nonextensivity parameter $q$) in
a Tsallis distribution to become complex. For completeness of the
presentation we start from the simple pure power law distribution,
\begin{equation}
O\left(x\right)=C\cdot x^{-m}. \label{eq:power}
\end{equation}
This function is scale invariant, i.e.,
\begin{equation}
O\left(\lambda x\right)=\mu O\left( x\right), \label{eq:sinv}
\end{equation}
with $m=-\ln{\mu}/\ln{\lambda}$. However, because $1 =
\exp{\left(\imath 2\pi k\right)}$, one can as well write that
\begin{equation}
\mu\lambda^{m} = 1 = \exp{\left(\imath 2\pi k\right)},\quad k = 0,
1, \dots . \label{eq:kkk}
\end{equation}
It means therefore that, in general,  the index $m$ can become
complex,
\begin{equation}
m=-\frac{\ln\mu}{\ln\lambda}+\imath\frac{2\pi k}{\ln\lambda}.
\end{equation}
As will be obvious from further, general considerations, such a form
of the power index results in $R$ as given by
Eq.~(\ref{eq:Factor}) when one only keeps $k = 0, 1$
terms (which is the usual assumption customary applied in all
applications \cite{Scaling,EQ,CHM,BD,RQFM}).

However, Tsallis distribution is only a power-like, not a power
distribution. Therefore, to explain the origin of such a dressing
factor in this case one has to find a right variable in which the real scaling holds.
We start from the observation that, whereas the Boltzmann-Gibbs (BG) distribution,
\begin{equation}
f(E) = \frac{1}{T} \exp\left( - \frac{E}{T}\right), \label{eq:BG}
\end{equation}
comes from the simple equation,
\begin{equation}
\frac{df(E)}{dE} = - \frac{1}{T} f(E), \label{eq:BGde}
\end{equation}
with the scale parameter $T$ being constant, the same equation,
but with variable scale parameter in the form
\begin{equation}
T = T(E) = T_0 + \frac{E}{n},  \label{eq:variableT}
\end{equation}
(known as {\it preferential attachment} in networks
\cite{NETWORKS,qWW1}\footnote{It is worth recalling here that
this very same form, $T(E) = T_0 + (1 - q)E$, also appears in
\cite{WalRaf} within a Fokker-Planck dynamics applied to the
thermalization of quarks in a quark-gluon plasma by collision
processes.} ),
\begin{equation}
\frac{df(E)}{dE} = - \frac{1}{T(E)}f(E) = - \frac{1}{T_0 +
E/n}f(E), \label{eq:net}
\end{equation}
results in the Tsallis distribution
\begin{equation}
f(E) = \frac{n - 1}{nT_0} \left( 1 + \frac{E}{nT_0}\right)^{-n}.
\label{eq:T}
\end{equation}
We shall write now Eq. (\ref{eq:net}) in finite difference form,
\begin{equation}
f(E + \delta E) = \frac{-n \delta E + nT + E}{nT + E} f(E).
\label{eq:netD}
\end{equation}
In practical sense this means a first-order Taylor expansion
for small $\delta E << E$ (from Eq. (\ref{eq:netD}) on, we use $T$
instead of $T_0$). We shall now consider a situation in which $\delta E$
always remains finite (albeit, depending on the value of the new
scale parameter $\alpha$, it can be very small) and equal to
\begin{equation}
\delta E = \alpha nT(E) = \alpha (nT + E).  \label{eq:dEalpha}
\end{equation}
Because one expects that changes $\delta E$ are of the order of
the temperature $T$, the scale parameter must be limited by $1/n$,
i.e., $\alpha < 1/n$. In this case, substituting
(\ref{eq:dEalpha}) into (\ref{eq:netD}), we have,
\begin{equation}
f[E + \alpha(nT + E)] =  (1 - \alpha n)f(E). \label{eq:scaling1}
\end{equation}
Expressing  Eq. (\ref{eq:scaling1}) in a new variable $x$,
\begin{equation}
x = 1 + \frac{E}{nT}, \label{eq:x}
\end{equation}
we recognize that the argument of the function on the left-hand
side of equality (\ref{eq:scaling1}) is
$$E + \alpha (nT + E) = (1 + \alpha)xnT - nT,$$
while the argument of the function on its right-hand side is
$$E = xnT - nT.$$
Notice that, in comparison with the right-hand side, the variable $x$
on the left-hand side is multiplied by the additional factor $(1 + \alpha)$. This
means that, formally, Eq.(\ref{eq:scaling1}), when expressed in
$x$, corresponds to the following scale invariant relation:
\begin{equation}
g[(1 + \alpha )x] = (1 - \alpha n)g(x). \label{eq:scaling}
\end{equation}
This means than that, following the discussion after Eq. (\ref{eq:sinv}),
its general solution is a power law,
\begin{equation}
 g(x) = x^{-m_k}, \label{eq:power}
 \end{equation}
 with exponent $m_k$ depending on $\alpha$ and acquiring an imaginary part,
 \begin{equation}
 m_k = - \frac{\ln ( 1 - \alpha n)}{\ln (1 +
 \alpha)} + ik \frac{2\pi}{\ln(1 + \alpha)}. \label{eq:solution}
\end{equation}
The special case of $k=0$, i.e., the usual real power law solution
with $m_0$ corresponding to fully continuous scale
invariance\footnote{In this case power law exponent $m_0$
still depends on $\alpha$ and increases with it roughly as $m_0
\simeq n + \frac{n}{2} (n+1)\alpha + \frac{n}{12}\left(4n^2 + 3n
-1\right)\alpha^2 + \frac{n}{24}\left( 6n^3 + 4n^2 - n
+1\right)\alpha^3 + \dots$. Notice also that $ \alpha < 1/n$. },
recovers in the limit $\alpha \rightarrow 0$ the power $n$ in the
usual Tsallis distribution. In general one has
\begin{equation}
g(x) = \sum_{k=0}w_k\cdot {\rm Re}\left( x^{-m_k}\right) = x^{-
{\rm Re}\left( m_k\right)}\sum_{k=0}w_k\cdot \cos\left[ {\rm
Im}\left(m_k\right) \ln(x) \right]. \label{eq:fin}
\end{equation}
One therefore obtains a Tsallis distribution decorated by a
weighted sum of log-oscillating factors (where $x$ is given by Eq.
(\ref{eq:x})). Because usually in practice we do not {\it a
priori} know the details of the dynamics of processes under
consideration (i.e., we do not known the weights $w_k$), for
fitting purposes one usually uses only $k=0$ and $k=1$. In this
case one has, approximately,
\begin{equation}
g(E) \simeq \left( 1 + \frac{E}{nT}\right)^{-m_0}\left\{ w_0 +
w_1\cos\left[ \frac{2\pi}{\ln (1 + \alpha)} \ln \left( 1 +
\frac{E}{nT}\right)\right]\right\} \label{eq:approx}
\end{equation}
and reproduces the general form of a dressing factor given by Eq.
(\ref{eq:Factor}) and often used in the literature \cite{Scaling}.
In this approximation the parameters $a$, $b$, $c$, $d$ and $f$
from Eq. (\ref{eq:Factor}) get the following meaning:
\begin{eqnarray}
\frac{a}{b} = \frac{w_0}{w_1},\qquad c = \frac{2\pi}{\ln (1 + \alpha)},\qquad
d = n T,\qquad f = - \frac{2\pi}{\ln(1 + \alpha)} \ln(nT). \label{eq:parameters}
\end{eqnarray}

In fact this is not the most general result for in our derivation,
Eqs.(\ref{eq:dEalpha})-(\ref{eq:scaling})), we have so far only accounted for a
single step evolution. In real situation one should expect to have a whole
hierarchy of evolutions. In such a case consecutive steps of evolution are connected
by:
\begin{equation}
E_i = E_{i-1} + \alpha_{i-1} \left( nT + E_{i-1} \right),
\label{eq:cascade}
\end{equation}
each with its own scale parameter $\alpha_i$. In the simplest
situation, neglecting any fluctuations of consecutive scaling
parameters, i.e., assuming that all $\alpha_i = \alpha$, one has
that after $\kappa$ steps
\begin{equation}
nT + E_{\kappa} = ( 1 + \alpha)^{\kappa} \left( nT + E_0 \right).
\label{eq:steps}
\end{equation}
This means that, in general, Eq. (\ref{eq:scaling}) should be replaced by
a new scale invariant equation:
\begin{equation}
g\left[ (1 + \alpha)^{\kappa} x\right] = ( 1 - \alpha n)^{\kappa}
g(x). \label{eq:scallingK}
\end{equation}
Whereas this equation does not change the slope parameter $m_0$,
it significantly influences the frequency of oscillations which
are now $\kappa$ times smaller,
\begin{equation}
c = \frac{2 \pi}{\kappa \ln (1 + \alpha)} \label{eq:kappa}
\end{equation}
(in Eq.(\ref{eq:scallingK}) $\lambda = (1 + \alpha)^{\kappa}$ and
$\mu = (1 - \alpha n)^{\kappa}$; the slope parameter $m_0 = - \ln
\mu/\ln \lambda$ is independent of $\kappa$, whereas the frequency
of oscillations, $2\pi/\ln \lambda$, decreases with $\kappa$ as
$1/\kappa$). For more complex behavior of intermediate scale
parameters $\alpha_i$ one gets more complicated expressions (we
shall not discuss this here).

\section{Other consequences of complex nonextensivity parameter}
\label{sec:III}

There are other consequences of allowing the parameter $m$ to be
complex. In what follows we shall discuss shortly three examples:
complex heat capacity , complex probability and complex multiplicative noise.

\subsection{Complex heat capacity}
\label{sec:CHC}

The complex power exponent in the Tsallis distribution, $m = m' + i\cdot m''$, means that
\begin{equation}
q -1 = \frac{1}{m} = \frac{m}{|m|^2}' + i\frac{m''}{|m|^2}. \label{eq:c_q1}
\end{equation}

As shown in \cite{BiroC} (cf. also \cite{Cq,qWW1,qWW}), the nonextensivity parameter
$q$ can be treated as a measure of the thermal bath heat capacity $C$ with
\begin{equation}
C = \frac{1}{q - 1} = m' + i m''.\label{eq:HC}
\end{equation}
The complex nonextensive parameter $q$ must therefore have some profound
consequences because now the corresponding heat capacity becomes
complex as well.  As a matter of fact, such complex (frequency
dependent) heat capacities (or generalized calorimetric
susceptibilities) are known in the literature \cite{HC} and are usually written in the form
\begin{equation}
C  = C_{\infty} + \frac{C_0 - C_{\infty}}{1 +  (\omega
\tau)^2}(1 - i \omega \tau).  \label{eq:hc}
\end{equation}
Here $C_{\infty}$ is the heat capacity related to the infinitely
fast degrees of freedom of the system as compared to the frequency
$\omega$, and $C_0$ is the total contribution at equilibrium (the
frequency is set to zero) of the degrees of freedom, fast and
slow, of the sample. The time constant $\tau$ is the kinetic
relaxation time constant of a certain internal degree of freedom.

These complex heat capacities are known as dynamic heat capacities
and are intensively explored from both experimental and
theoretical perspectives. It is expected that dynamic calorimetry
can provide an insight into the energy landscape dynamics, cf.,
for example, \cite{G2007,GR2007,ND,S}. Usually one associates the
imaginary part of linear susceptibility with the absorption of
energy by the sample from the applied field.

In the case of temperature fluctuations  $ \delta T(t)$ the
deviation of the energy from its equilibrium value  $\delta U(t)$
is, for a certain linear operator $\hat{C}(t)$, some linear
function of the corresponding variation of the temperature,
\begin{equation}
\delta U(t) = \hat{C} \delta T(t). \label{eq:hatC}
\end{equation}
If the temperature of the reservoir changes infinitely slowly in
time, then the system can keep up with any changes in the
reservoir and its susceptibility is just the specific heat of the
system $C_V$. However, in general, the behavior of the system is
described by a generalized susceptibility $C_V(\omega)$, which can
be called {\it the complex and $\omega$-dependent} heat capacity
of the system. The change in the energy of a
system in the field of the thermal force can be represented by
\begin{equation}
\delta U(t) = \int L\left(t'\right)\delta T\left(t -
t'\right)dt', \label{UL}
\end{equation}
where $L\left( t'\right)$ is the response function
of the system describing its relaxation properties given by
$\Phi(t) = \int_t^{\infty}L\left( t'\right)dt'$. Taking the
Fourier transform one gets
\begin{equation}
\delta U(\omega) = C_V(\omega)\delta T(\omega), \label{deltaU}
\end{equation}
where
\begin{equation}
C_V(\omega) = \int L\left( t'\right)e^{i\omega t'} dt' \label{CVOM}
\end{equation}
is the generalized susceptibility of the system and is
called the complex heat capacity. In practice, the frequency
dependent heat capacity is a linear susceptibility describing the
response of the system to the small thermal perturbation
(occurring on the time scale $1/\omega$)  that takes the system
slightly away from the equilibrium .

A complex $C_V(\omega)$  means that $\delta U$  and $\delta T$ are
shifted in phase and that the entropy production in the system
differs from zero \cite{S}. The corresponding
fluctuation-dissipation theorem for the frequency dependent heat
capacity was established in \cite{ND}. According to this result,
the frequency-dependent heat capacity may be expressed within the
linear response approximation as a linear susceptibility
describing the response of the system to arbitrarily small
temperature perturbations away from equilibrium,
\begin{equation}
C_V(\omega) = \frac{\langle U^2\rangle_0}{\langle T\rangle^2}\,  -\,  i\frac{\omega}{\langle T\rangle^2}
 \int_0^{\infty} dt e^{-i\omega t} \langle U(0) U(t)\rangle \label{eq:C_V}
\end{equation}
(the $\omega$ denotes frequency with which temperature field is
varying with time).

The above results for heat capacity can now be used to a new
phenomenological interpretation of the complex $q$ parameter
discussed before. Namely, one can argue that
\begin{equation}
q - 1 = \frac{Var(T)}{\langle T\rangle^2} - i\, \frac{S(T)}{\langle
T\rangle^2}, \label{eq:qCq}
\end{equation}
were
\begin{equation}
S(T) = \omega \int  Cov[ T(0),T(t) ] e^{-i\omega t}\,
dt \label{eq:ST}
\end{equation}
is the spectral density of temperature fluctuations (i.e., the
Fourier transform of the covariance function averaging over the
nonequilibrium density matrix).

We would like to stress at this point that, in a sense, Eq.
(\ref{eq:qCq}) can be regarded as a generalization of our old
proposition for interpreting $q$ as a measure of nonstatistical
intrinsic fluctuations in the system \cite{WWB,SSB} (which
corresponds to the real part of (\ref{eq:qCq})) by adding the
effect of spectral density of such fluctuations (via the imaginary
part of (\ref{eq:qCq})). Notice that (\ref{eq:qCq}) follows from
(\ref{eq:HC}) and the relation $U=C_V T$, allowing to write
(\ref{eq:C_V}) in the form of (\ref{eq:qCq}).

\subsection{Complex probability}
\label{eq:CP}

From the point of view of superstatistics \cite{SuperS,WWB}, in
our particular case  complex parameter $q$ corresponds to a
complex probability distribution. Namely, one uses the property
that gamma-like fluctuation of the scale parameter $T$ in an
exponential BG distribution (\ref{eq:BG}) results in the
$q$-exponential Tsallis distribution (1) with $q > 1$. The
parameter $q$ is given here by the strength of these fluctuations,
$q = 1 + Var(X)/<X>^2$.  From the thermal perspective, it
corresponds to situation in which the heath bath is not
homogeneous but has different temperatures in different parts,
which are fluctuating around some mean temperature $T_0$. It must
be therefore described by two parameters: a mean temperature $T_0$
and the mean strength of fluctuations given by $q$.

We now perform the same procedure, but using two gamma distributions, one with
a real power index, $m_0 - 1$, and one with a complex power index, $m_0 + i m_1 -1$,
\begin{eqnarray}
g(1/T) &=& w_0\frac{1}{\Gamma\left(m_0\right)} nT_0\left( n\frac{T_0}{T}\right)^{m_0 -1 }
\exp\left( - n\frac{T_o}{T}\right) +\nonumber\\
       && + w_1 \frac{1}{\Gamma\left(m_0 + i m_1\right)} nT_0
\left( n\frac{T_0}{T}\right)^{m_0 + i m_1 -1} \exp\left( - n\frac{T_0}{T}\right).\label{DGamma}
\end{eqnarray}
As the result one gets a complex distribution (complex pdf):
\begin{equation}
h_q(E) = \int_0^{\infty}\, f(E) g(1/T) d(1/T) = C w_0\left(1 + \frac{E}{nT_0}\right)^{-m_0} +
C w_1 \left( 1 + \frac{E}{nT_0}\right)^{-m_0 - i m_1}, \label{hq}
\end{equation}
the real part of which is pdf in form of a Tsallis distribution decorated with log-periodic oscillations of the type
of Eq. (\ref{eq:approx}),
\begin{equation}
Re\left[ h_q(E)\right] = C\left( 1 + \frac{E}{nT_0}\right)^{-m_0}\cdot \left\{ w_0
+ w_1\cos \left[ m_1 \ln \left(1 + \frac{E}{nT_0}\right)\right]\right\}. \label{Complexh}
\end{equation}

The complex pdf has a number of interesting properties \cite{HB,MZ}. It plays an important role in
the interference among resonance states during scattering experiments.
It is associated with the phase of the resonance channel probability amplitudes
(in non-Hermitian quantum mechanics).  In wireless communication systems it is
generated by a superposition of finite random variables and usually involves the movement,
scattering, diffusion or diffraction. The imaginary part is proportional to the degree of
the correlation. The imaginary part is then a function of a correlation coefficient or
other parameters that state the degree of the relationship of each individual random
variable of the superposition of the random variable having a complex pdf.
The real and imaginary part have diverse properties, i.e. one for real valued pdf
and the other for elementary correlation, respectively.

It is interesting to note that entropy
\begin{equation}
H = - \left| \int \int ( a\ln a + i\cdot b\ln b) dx_1 dx_2 \right|
=
\end{equation}
corresponding to complex joint probability,
\begin{equation}
f\left(x_1,x_2\right) = a\left(x_1,x_2\right) + i\cdot
b\left(x_1,x_2\right), \label{cjp}
\end{equation}
consists of two components:
\begin{equation}
\!\! H_1 = - \int\!\! \int a \ln a\, dx_1 dx_2,\quad H_2 = - \int \!\! \int b \ln b\, dx_1 dx_2;
\quad H = \left| H_1 + i H_2\right|\sqrt{H_1^2 + H_2^2} \ge H_1. \label{EH12}
\end{equation}
The imaginary part of entropy is proportional to the degree of incompatibility of the
correlated stochastic processes. The incompatibility increases the entropy of correlated
stochastic processes.

\subsection{Complex multiplicative noise}
\label{sec:CMN}

It is known that multiplicative noise leads to a Tsallis
distribution \cite{SSB}. It is then natural to expect that
multiplicative complex noise should result in complex $q$ and in
log-periodic oscillations in Tsallis distributions. It can be
defined by a Langevin equation
\begin{equation}\frac{dp}{dt} +\gamma(t)p = \xi(t),\quad{\rm where}\quad \gamma(t) =
\gamma_0(t) + i\gamma_1. \label{Cl}
\end{equation}
The resulting distribution \cite{SSB} is now
\begin{equation}
f(p) = \left(1 + \frac{q - 1}{T}p^2 \right)^{\frac{q}{q-1}}\quad
{\rm where}\quad T = \frac{2Var(\xi)}{\langle \gamma\rangle},\quad
q = 1 + \frac{2Var(\gamma)}{\langle \gamma\rangle}. \label{fpc}
\end{equation}
The parameter $q$ is now complex because $\langle \gamma\rangle$
is complex. Even more importantly, $(q - 1)/T =
Var(\gamma)/Var(\xi)$ is real (it tends to zero for $ q
\rightarrow 1$). This is because the complex term $\gamma_1$ added
to the noise is constant.  Notice that we could just as well
replace in Eq. (\ref{fpc}) $(q - 1)\left(p^2/T\right)$ by $\left(
p^2/p_0^2\right)$ where $p_0^2 = Var(\xi)/Var(\gamma)$. The
examples and discussion of the systems characterized by the
appearance of "imaginary" multiplicative noise terms in an
effective Langevin-type description can be found in
\cite{CLang}\footnote{In fact, this is not exactly Tsallis formula
from Eq. (\ref{T}). To get it one has to allow for correlation
between noises and drift term due to additive noise, i.e., for
$Cov(\xi,\gamma) \neq 0$ and $\langle \xi \rangle \neq 0$ (see
\cite{RodosWW} for details). One obtains then Eq. (\ref{T}) but
with, in general, complex  $T=T(q)$. We shall not discuss it
here.}.

\section{Summary and conclusions}
\label{sec:IV}

In may places in physics, and especially in the realm of high
energy multiparticle production processes we are particularly
interested in, it became a standard procedure to fit the data on
transverse momentum distributions by means of the quasi-power
Tsallis formula. The usual interpretation in such cases is that
the scale parameter $T$ is a kind of "temperature" whereas
additional nonextensivity parameter $q$ is describes intrinsic,
nonstatistical fluctuations existing in the system
\cite{BCM,Beck,RWW,qWW1,qWW,Wibig,Biro,BiroC,JCleymans,ADeppman,Others,WalRaf,qCR,SuperS,WWB,SSB,RodosWW}.
However, with increasing range of transverse momenta measured in
recent experiments \cite{CMS,ATLAS,ALICE} two things happened:
\begin{itemize}
\item[$(i)$] That they still can be fitted  by the same formula
(which came as surprise because fits now cover $\sim 14$ orders of magnitude of the
measured cross sections \cite{CYW,ISMD2014}).
\item[$(ii)$] That new data revealed weak but persistent oscillation
of log-periodic character (discussed already shortly in \cite{LPOWW}).
\end{itemize}
If taken seriously, such log-periodic structures in the data indicate that
the system and/or the underlying physical mechanisms have
characteristic scale invariant behavior. This is interesting as
it provides important constraints on the underlying physics. The
presence of log-periodic features signals the existence of
important physical structures hidden in the fully scale invariant
description. It is important to recognize that Eq. (\ref{eq:net})
represents an averaging over highly 'non-smooth' processes and, in
its present form, suggests rather smooth behavior. In reality,
there is a discrete time evolution for the number of steps. To
account for this fact, one replaces a differential  Eq.
(\ref{eq:BGde}) by a difference quotient and expresses $dt$ as a
discrete step approximation given by Eq. (\ref{eq:dEalpha}) with
parameter $\alpha$ being a characteristic scale ratio. It can also
be shown that discrete scale invariance and its associated complex
exponents can appear spontaneously, without a pre-existing
hierarchical structure. Finally, a complex nonextensivity parameter
promises new perspectives in future phenomenological applications
being connected to complex heat capacity, to notion of complex
probability or to complex multiplicative noise, to mention only a
few examples discussed shortly in our paper.


\acknowledgments{Acknowledgments}

This research was supported in part by the National Science Center
(NCN) under contract DEC-2013/09/B/ST2/02897. We would like to
warmly thank Dr Eryk Infeld for reading this manuscript.


\authorcontributions{Author Contributions}

The content of this article was presented by Z. W\l odarczyk at the
Sigma Phi 2014 conference at Rhodes, Greece.


\conflictofinterests{Conflicts of Interest}

The authors declare no conflict of interest.

\bibliographystyle{mdpi}
\makeatletter
\renewcommand\@biblabel[1]{#1. }
\makeatother

\end{document}